\documentclass[twocolumn,showpacs,preprintnumbers,amsmath,amssymb]{revtex4}
\usepackage{graphicx}
\usepackage{dcolumn}
\usepackage{bm}
\usepackage{subfigure}
         
\newcommand{\gam}{\ensuremath{\gamma} }
\newcommand{\dgam}{\ensuremath{\dot{\gamma}} }
\newcommand{\s}{\ensuremath{\sigma} }
\newcommand{\ie}{{\it i.e. }}
\newcommand{\etal}{{\it et al. }}

\begin{document}
\title{A simple model for heterogeneous flows of yield stress fluids}
\author{Guillemette Picard}
\email{guillemette@turner.pct.espci.fr}

\author{Armand Ajdari}
\affiliation{Laboratoire de Physico-Chimie Th\'eorique, UMR CNRS 7083, ESPCI, 10 rue Vauquelin, F-75231 Paris Cedex 05, France}
\author{Lyd\'eric Bocquet}
\affiliation{Laboratoire de Physique de l'ENS de Lyon, UMR CNRS 5672, 46 all\'ee d'Italie, 69364 Lyon Cedex, France}
\author{Fran\c{c}ois Lequeux}
\affiliation{Laboratoire de Physico-Chimie Macromol\'eculaire, UMR CNRS 7615, ESPCI, 10 rue Vauquelin, F-75231 Paris Cedex 05, France}

\date{Revised manuscript sent to PRE, August 30 2002}
  
\begin{abstract}
Various experiments evidence spatial heterogeneities in sheared yield stress fluids. To account for heterogeneities in the velocity gradient direction, we use a simple model corresponding to a non-monotonous local
 flow curve and study a simple shear geometry. Different types of boundary conditions are considered. 
Under controlled macroscopic shear stress $\Sigma$, we find homogeneous flow in the bulk and a hysteretic macroscopic stress - shear rate curve. 
Under controlled  macroscopic shear rate $\dot{\Gamma}$, shear banding is predicted within a range of values of $\dot{\Gamma}$. For small shear rates, stick slip can also be observed. 
These qualitative behaviours are robust to changes in the boundary conditions.
 \end{abstract}
\maketitle

%%%%%%%%%%%%%%%%%%%%%%%%%%%%%%%%%%%%%%%%%%%
\section{Introduction}
%%%%%%%%%%%%%%%%%%%%%%%%%%%%%%%%%%%%%%%%%%%
\label{introduction}

Complex materials, such as concentrated colloidal suspensions, gels or foams, often exhibit 
exotic rheological behaviour at low shear \cite{britt,ram,berret1,diat,deb,che,cous,sol,pig,cloitre,bar}, very different from the
Newtonian linear response characterised by a stress $\Sigma$ linearly proportional to the shear rate
$\dot \Gamma$ (in a scalar language).
A term often coined for some of these behaviours is that of ``yield stress fluids''.
Although often loosely used, it most of the time describes fluids that have one or both 
of the following properties: when one looks at the steady state 
flow curve $\Sigma(\dot \Gamma)$ at low shear, 
i) the stress $\Sigma$ tends to a finite value $\Sigma_Y$ when
the shear rate goes to zero, ii) no flowing steady state solution exists for stresses smaller
than a value $\Sigma_0$. Consequently at low stresses, the system does not really flow,
and these ``pastes'' often display physical aging in a way rather similar to structural glasses.  

\begin{figure}[t]
\includegraphics[width=7.5cm]{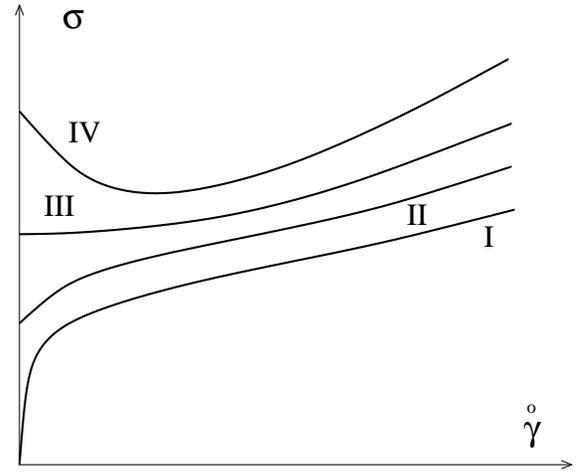}
\vspace{.5cm}
\caption{Schematic representation of
possible local steady state curve $\sigma(\dot \gamma)$
generating a macroscopic yield-stress fluid behaviour.}
\label{gen}
\end{figure}

A traditional way to apprehend this {\em macroscopic} rheological 
non-linear behaviour is to suppose the existence of a {\em local}
relation between the {\em local} stress $\sigma$ and the {\em local} shear rate $\dot \gamma$
of one of the forms I to IV on figure \ref{gen}. Classically again, I to III being monotonic
can lead to a homogeneous flow so that the macroscopic response
is identical to the local flow curve : 
while II and III correspond to genuine yield stress fluids 
satisfying both i) and ii) above (with $\Sigma_0=\Sigma_Y=\sigma ({\dot\gamma=0})$ ),
I yields a power law fluid behaviour
$\Sigma = A {\dot \Gamma}^{\alpha}$ which can be experimentally very difficult to tell
apart from the yield stress behaviours II and III if $\alpha$ is small.
It is on the other hand classically argued that the decreasing branch of IV
 ($d\Sigma/d{\dot \Gamma} <0$) can not lead to the corresponding  homogeneous
situation, as such a flow is unstable with respect to the formation of ``shear bands''
or domains flowing at different {\em local} shear rates. This statement
is usually substantiated by a simple argument invoking inertial terms (see, for instance \cite{Dhont}). 
This type of banding is referred to as 'gradient banding' by opposition to 'vorticity banding'
(coexistence of domains flowing at different {\em local} shear stress and common shear rate).

The question then arises of the structure of the heterogeneous flow pattern observed 
in these conditions (curve IV, gradient banding), 
of its stability (i.e. are there always steady-state solutions?),
and of the resulting macroscopic flow curve $\Sigma({\dot\Gamma})$ 
for the steady-state solutions.
Investigating this problem in generic terms is very difficult as
many underlying dynamical laws involving an unlimited number of ``hidden'' variables
can generate local equations such as IV.
In this paper, to make progress, we consequently make a specific choice for this underlying 
dynamics, that we justify now. \\

Formally, taking the vertical $\dot \gamma =0$ axis  as part of the local
flow curve,
the problem is very similar to the analysis of flow-induced
or flow-assisted ``phase'' transitions for systems with a local non-monotonic relationship $\sigma(\dot \gamma)$.
For some systems,
the change in rheological properties can be ascribed to a structural phase transition occurring 
under flow : Ramos \etal with lyotropic hexagonal phases  \cite{ram}, Diat \etal with onions \cite{diat} and 
Britton \etal and Berret \etal with wormlike micelles \cite{britt,berret1} observe a phase transition, and the coexistence of two
well defined phases. Such behaviours are quite
reminiscent of equilibrium phase transitions, although no general principle applies {\it a priori}
in such out of equilibrium situations \cite{Harrowel}.
The identification of structural differences between the flowing phases then
provides a guide as to what are the most likely relevant local variables beyond the rheological ones:
a (non-conserved) nematic order parameter, the (conserved) concentration of a constituent, etc... 
Related theoretical descriptions have thus tried 
to describe the coupling between this order parameter and flow in order to 
obtain a selection criterion for the flow pattern and the resulting macroscopic rheological behaviour \cite{Nozieres,olm0,olm3,gov,olm1,olm2,Harrowel}.\\

However for many yield stress fluids there is no obvious structural
difference between the coexisting
flowing phases: experimental reports often describe "shear bands" of a flowing 
"phase"  in a stationary "phase" with no clear structural or textural difference 
between the two parts of the system. 
Such ``soft glassy 
materials" \cite{sol} include gels, pasty systems, concentrated glassy suspensions, \dots 
One may cite the work by Raynaud \etal \cite{cous},
where banding is observed in a suspension of bentonite and the interface between 
the bands is in the velocity-vorticity plane (gradient banding) : the system is
sheared within a band at the inner moving cylinder, while the rest of the sample 
remains un-sheared. Laponite suspensions also undergo banding 
for a certain range of imposed shear rates, range for which the shear stress 
is almost constant \cite{pig}. Noticeably, for a range a lower shear rates 'stick slip' is observed 
in this system : a layer reversibly fractures and re-heals, showing stationary oscillations 
in the measured stress.
Eventually, one may note that
inhomogeneous velocity profiles are also observed in sheared foams \cite{deb}, and granular
pastes \cite{bar}.

For these systems with no clearly identified structural variable 
that would distinguish coexisting phases, there is much less guidance for building models. 
Still, banding  is likely to originate from the same basic physical mechanism 
as for the 'phase separating' fluids \cite{porte}. 
Models have therefore been derived along similar lines, 
which differ by the choice of the hidden variable (equivalent to the order parameter) 
and of its dynamics \cite{porte,pearson}. 
In addition or alternatively, gradient terms were included in the 
non monotonic intrinsic law $\s(\dgam)$ \cite{Dhont,Lu}. 

In the present paper we follow the route
initiated by Derec \etal \cite{car}, that has been shown to be a simple though useful guide
to understand the interplay between non-linear rheology and aging properties
for yield-stress fluids characterised by local flow curves
of types I to III \cite{car,carthese}. 
In the present work we adapt it to describe local flow curves
of type IV, and to take into account spatial heterogeneities.

The essence of the model of \cite{car} is to describe the local state of
the system by a single scalar variable that is for convenience taken to be the local
relaxation frequency of the stress (called ``fluidity''), as expressed by a Maxwell equation: 
$\partial_t \sigma =-a(t) \sigma +{\dot \gamma}$ (stress has been rescaled so that
the elastic modulus is $1$).
Then the evolution of the fluidity $a(t)$ results from the competition between:
(i) {\it aging - } relaxation rates in the system decrease during 
the spontaneous evolution of the pasty system,
this evolution being itself  slower and slower as time passes; 
and (ii) {\it flow-induced rejuvenation -},
flow induces rearrangements in the system, which in turn increase the "fluidity" of the system
as a feed back mechanism \cite{cloitre,carthese,kurchan,bonn}.  This competition results in an equation
for $a$ that supplements the Maxwell one: $\partial_t a= -f(a) +g(a,\sigma,\dot \gamma)$,
with $f$ and $g$ positive functions.

Here, to address the question at hand,
 we focus on choices for $f$ and $g$ leading to local flow curves 
(local steady-states)  with a decreasing branch at low values of $\dot \gamma$, 
and we attempt to describe spatial heterogeneities,
through a simple diffusion term $  D \Delta a$  in the equation describing the evolution of $a$.

We then study the flow patterns in a simple shear geometry for such a
system, when it is driven at either imposed macroscopic shear stress $\Sigma$
or imposed macroscopic shear rate $\dot \Gamma$, and deduce the resulting
macroscopic rheological behaviour.
We will on the way provide analytical and numerical arguments,
using when necessary explicit choices for $f$ and $g$,
and pinpoint the role of the boundary conditions.
 In particular we will show that
the model accounts for the following features :
\begin{itemize}
\item when $\Sigma$ is imposed, the flow curve jumps in a hysteretic 
fashion between two branches: one that corresponds to no flow
but for a wall layer in the vicinity of the walls;
the other corresponds to a fully fluidised situation. 
\item when the global shear rate $\dot\Gamma$ 
is imposed, 
shear banding can occur, 
as well as sometimes
a stick-slip like oscillating behaviour at small shear rate that corresponds 
to a localised oscillation
of the fluidity.
\end{itemize}
These features will be analysed in the context of experimental results.

The paper is organised as follows.
We recall in Section \ref{model} the basic ingredients of the model in Ref. \cite{car} and propose
a generalisation to account for inhomogeneities in the system.
The homogeneous case is detailed in section \ref{homogeneous} and the stability 
of the homogeneous solutions
is discussed. 
Heterogeneous solutions are considered in Sections \ref{stress_H} and \ref{shearrate}.
In section \ref{stress_H}, the controlled stress situation is considered, 
while in section \ref{shearrate}, the global shear rate is imposed.
In section \ref{discussion}, choices made in building the model and the influence of 
the boundary conditions are discussed.

%%%%%%%%%%%%%%%%%%%%%%%%%%%%%%%%%%%%%%%%%%%%%%%%%%
\section{Model and geometry}
\label{model}
%%%%%%%%%%%%%%%%%%%%%%%%%%%%%%%%%%%%%%%%%%%%%%%%%%%%

In this section we define more clearly the model and the geometry we are going to study, following the physical arguments given in the introduction. In subsection \ref{sub1}, we specify further
the functions $f$ and $g$, in subsection \ref{sub2} we describe the geometry considered and the corresponding
description of heterogeneities in the equations, in subsection \ref{sub3} we define the boundary conditions for
both velocity and fluidity, and we summarise in subsection \ref{sumup} by recalling the resulting set of 
equations.

%%%%%%
\subsection{Accounting for a non-monotonous local flow curve}

%%%%%%%%
\label{sub1}
From the introduction, we start with the model of \cite{car},
devised for homogeneous situations:
\begin{equation}
\left\{ 
\begin{array}{ll}
\partial_t \sigma = -a \sigma + \dot \gamma\\
\partial_t a = - f(a) + g(a,\sigma, {\dot \gamma}) 
\end{array} 
\right.
\label{eq_fluidity}
\end{equation}

For sake of explicitness, and at loss of generality, it will often be 
useful to restrict the choices for the functions $f$ and $g$ appearing in
equation (1). This is in particular true for the flow induced rejuvenation term
$g(a,\sigma,\dot \gamma)$ which can take many forms.

Following again Derec et al. \cite{car}, we take forms corresponding to low values of the stress, shear rate and fluidity, so that power laws may express a kind of expansion of more
general functions. Also $a$ being a positive scalar, 
the direction of shear is of no influence so one expects $g(a,\sigma,\dot\gamma)=
g (a,-\sigma,-\dot\gamma)$. A simple solution is then
$g(a,\sigma,\dot\gamma)=h(a) \sigma\dot\gamma$.

\begin{figure}
\includegraphics[width=7.5cm]{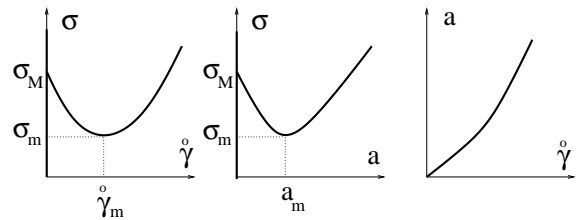}
\caption{Schematic drawing of the local steady state flow curve $\sigma(\dgam)$, $\s(a)$,$\dgam(a)$.}
\label{cur}
\end{figure}

Then following the introduction, we want the
steady-state flow curve $\sigma(\dot \gamma)$
to have a decreasing branch for small but non-zero values of $\dot \gamma$ 
(as sketched in figure \ref{cur}). 
In that case, the steady-state solutions of the system (\ref{eq_fluidity}), consist of :
the vertical axis (
$a=0$, $\dot \gamma =0$, $\sigma$ arbitrary),
and solutions given by $\sigma(a)=(f(a)/ah(a))^{1/2}$ when expressed in the plane ${s,a}$.
The function $a(\dgam)$ is a monovalued increasing function, whereas $\sigma(a)$ is multivalued,
decreasing for small values of $a$, then increasing for larger values. The requirements are thus that that $f(a)/ah(a)$ has a negative derivative for $0<a<a_m$,
a positive one for $a_m <a$. The minimum thus reached for $a_m$ corresponds
to the minimal value of the stress $\sigma_m=(f(a_m)/a_mh(a_m))^{1/2}$.
Eventually, as for curve IV in Figure 1, 
$\sigma$ tends towards a finite value when $\dot \gamma$ goes to zero,
which we call  $\sigma_M$.

When numerical calculation will prove necessary to analyse the evolution of the system, we 
make even more explicit choices regarding $f$ and $h$, that need of course satisfy 
the above requirements. 
We use the simple choice: $f(a)= r a^2 + a^4$, $h(a)=r' a + a^2$ with $r$ and $r'$ arbitrary
number that we vary to check the robustness of our conclusions. That $f(a)$ scales
as $a^2$ for small values
of $a$ gives a spontaneous relaxation of $a$ as the inverse of the elapsed time, a situation often
referred to as ``full aging'' (see \cite{car}).

%%%%%%
\subsection{Incorporating heterogeneities: geometry and model}
%%%%%%
\label{sub2}
In principle, a  general description of
heterogeneous flows implies complex algebra
 describing the evolution of scalar and tensorial quantities through convected derivatives, 
and writing various conservation equations (for mass, momentum, etc ..).
We focus here on the simplest flow geometry (simple shear),
allowing only variations along one direction, perpendicular 
to flow and vorticity.

More precisely, we consider our complex fluid to be bounded 
by two parallel plates located at $z=0$ and $z=H$ (see e.g. Figure
3). The top plate moves along $x$ at a velocity $V$,
while the bottom one is immobile, so that the macroscopic shear rate is $\dot \Gamma= V/H$.
In this situation we consider laminar flows along $x$, where the local variables $\sigma=\sigma_{xz}$, 
$\dot \gamma$, and $a$, depend only  on $z$ (i.e. not on $x$ and $y$). 

\begin{figure}
\includegraphics[width=7.5cm]{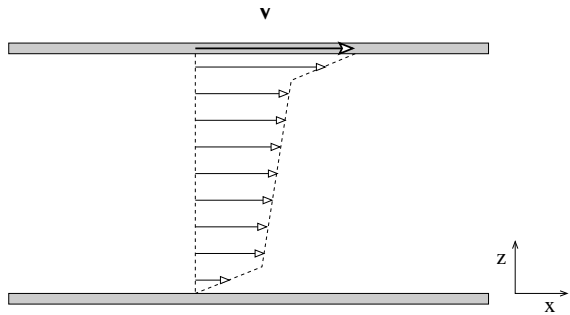}
\caption{Geometry: simple shear along $x$, variations along $z$ only.
The flow can be heterogeneous, the example depicted here correspond to
a low fluidity in the bulk and wall layers of higher fluidity
close to the wall.}
\label{ge0}
\end{figure}

Neglecting acceleration (inertial) terms is a common
assumption for such (highly viscous) fluids on the ground 
of a low Reynolds number.
This in general makes the stress divergence free, 
but in the present geometry it results in the stress being spatially constant
so that it is no more a local variable. We will in this paper make this assumption
of fast momentum diffusion, while keeping in mind
that by doing so we formally suppress a ``mechanical instability'' 
that destabilises into bands any homogeneous flow solution corresponding
to a decreasing part of the local flow curve $d\sigma/d\dot \gamma <0$ (see e.g.\cite{sch,Dhont,porte}). 
In this framework, the local stress depends solely on time,
and is identical to the macroscopic stress $\sigma(z,t) = \sigma(t)=\Sigma(t)$.

Now in order to discuss spatial heterogeneities, we need to account for the effect
on the dynamics of a heterogeneous value of the fluidity. 
Again for the sake of simplicity, we model this as
a diffusion process in the evolution of the fluidity and correspondingly add a $ D \nabla^2 a $ 
diffusion term to the second equation of (1). 
This choice corresponds to a picture in which the system is fluidised in a localised region, 
this triggers spontaneous local relaxations 
which themselves tend to fluidise the neighbourhood of the initial region.
Accordingly we take the ``diffusive coefficient'' to be a positive constant. 
A negative value would generate local, small wavelength instabilities that destabilise 
homogeneous flow even at high shear rates. As we focus here on the instability 
generated by the non monotonous branch of the local flow curve, 
and in line with experimental observations \cite{cous,cloitre} 
of homogeneous flow at high shear, we discard this possibility here. 
In addition, we do not consider other terms formally allowed by symmetry such as $ (\nabla a)^2$.

The system is therefore described by the coupled equations:

\begin{equation}
\left\{ 
\begin{array}{ll}
\partial_t \sigma(t) = -a(z,t) \sigma(t) + \dot \gamma(z,t)\\
\partial_t a(z,t) = - f(a(z,t)) + h(a(z,t)) \sigma(t) {\dot \gamma}(z,t) + \\
\mbox{\hspace{2cm}} D \partial^2_{zz}a(z,t)
\end{array} 
\right.
\label{eq_fluidity_d}
\end{equation}

One may note that in its present form, the equation for the fluidity is close to the equation
for the granular temperature obtained within granular hydrodynamics of inelastic
beads \cite{boc} : the spatial diffusion term accounts in the latter case for "heat" diffusion
(according to Fourier law) and the "aging" term corresponds to the energy dissipation term
(due to the inelasticity of collisions), which drives the granular material towards a state
of zero temperature. Eventually, the granular temperature is indeed 
a measure of the local fluidity of the granular material.

An important comment is also that the diffusion coefficient $D$ that we took constant here could
in principle be a function of $a$. This would affect the length scale appearing
from the competition of $f(a)$ and $D\partial^2_{zz}a$.
If as in the  example at the end of the previous subsection $f(a) \sim a^2$ for small $a$,
then this length scale depends on the actual value of $a$ : $(D/a)^{1/2}$, 
indicating algebraic tails for spatial structures of low fluidity.

%%%%%%%%%%%%
\subsection{Boundary conditions}
%%%%%%%%%%%%
\label{sub3}
The equation for the fluidity and velocity have to be supplemented by boundary conditions
at the confining walls at $z=0$ and $z=H$.

We here exclude any direct wall slippage  and assume the plates to be sufficiently 
rough for the plate velocity to be equal to the limit velocity in the fluid
approaching the plate (no slip velocity). This is equivalent to:
\begin{equation}
\int_0^Hdz \, \dot\gamma(z ,t)=V(t)= \dot \Gamma (t) H
\end{equation}
It is important to point out that this hydrodynamic condition is not sufficient : 
a boundary condition for the fluidity equation is also required.
How the walls affect the density, order, relaxation processes of the fluid 
in their vicinity is obviously a complex problem with many aspects to it, 
even for simple liquids \cite{boc2}. 
A smooth flat wall can enhance ordering, a rough or chemically 
disordered one can on the contrary locally disorganise the fluid (relative to what happens in the bulk). 
The consequences of these on the local relaxation rates being far from trivial \cite{sch1}, 
we choose here to investigate two limiting simple cases that hopefully bracket the true behaviour :
(i) a fixed fluidity at the wall 
\begin{equation}
a(z=0) = a(z=H)=a_w 
\end{equation}
and (ii) a zero gradient condition
\begin{equation}
\partial_{z}a|_{z=0,H}=0.
\end{equation}

%%%%%%%%%%%%%%%%%%%%%
\subsection{Summary}
%%%%%%%%%%%%%%%%%%%%%%
\label{sumup}
We summarise here the equations that we will use in the remainder of the paper.

\begin{equation}
\left\{ 
\begin{array}{ll}
\partial_t \Sigma = -a \Sigma  + \dot \gamma \\
\partial_t a = - f(a) + h(a) \Sigma {\dot \gamma} + D \partial^2_{zz}a\\
\int_0^Hdz \, \dot\gamma(z,t)=V= \dot \Gamma H \\
a|_{z=0,H}=a_w \,\,\, {\rm or}\,\,\, \partial_z a|_{z=0,H}=0  
\end{array} 
\right.
\label{eq_fluidity_d2}
\end{equation}

where $\Sigma(t)$, $\dot \Gamma(t)$ are global variables whereas $a(z,t)$ and $\dot\gamma(z,t)$
are local ones.
When explicit choices will be necessary to study the corresponding dynamics,
we will use $f(a)=ra^2 + a^4$ and $h(a)= r'a + a^2$.\\
Let us recall that by neglecting the inertial terms, 
we formally suppress the  ``mechanical instability'' 
that destabilises the decreasing branch $d\sigma / d \gam <0$.\\ 

{\em Numerical method -} For various cases, it is difficult/impossible to 
get analytical results and numerical calculation becomes necessary.
The evolution equations for the stress and the fluidity, Eq. (\ref{eq_fluidity_d2}), have been
solved numerically using a pseudo-spectral method, together with a 
Runge-Kutta integrator  \cite{spec}. This technique
integrates the diffusion part of the time equation for the fluidity in Fourier space, while
the non linear part is solved in real space. Special care has been taken to
avoid aliasing of the solutions. A standard Fast Fourier Transform (FFT)
algorithm is used for the forward and backward transformations between real
and Fourier space. \\

We wish to use equations (6) to deduce the macroscopic behaviour of the system,
to analyse the conditions for the existence of a steady-state, 
and to describe the features of the flow. Note that the 
boundaries appear explicitely, as potential heterogeneities, so that most of the time
the structure close to the boundaries will be different than in the bulk,
allowing the walls to nucleate new ``phases'' before the bulk.
Actually the absence of noise terms in these equations 
suppresses the equivalent of homogeneous nucleation in thermodynamics problems. 
We briefly comment upon the addition of noise when it plays a significant role.
In the following three sections, we first revisit briefly the formal case of a homogeneous behaviour
(section III),
we then analyse the simpler case of the response to a fixed stress $\Sigma$ (section IV),
before turning to the more complex case of a fixed global shear rate $\dot \Gamma$ (section V).

%%%%%%%%%%%%%%%%%%%%%%%%%%%%%%%%%%%%%%%%%%%%%%%%%%%%
\section{Homogeneous behaviour}
\label{homogeneous}
%%%%%%%%%%%%%%%%%%%%%%%%%%%%%%%%%%%%%%%%%%%%%%%%%%

We first analyse briefly the homogeneous case, where 
we impose that the fluidity $a(t)$ is 
independent of $z$, so that $\Sigma$ and $\dot \Gamma$ 
are similar to $\sigma$ and $\dot\gamma$. Although slightly redundant with previous discussions of
the flow curve, this analysis will highlight the potential differences
between the stability analyses at fixed stress and fixed shear rate. These differences
naturally result from the fact that $\Sigma$ and $\dot\Gamma$ are treated on a
different footing in equation (\ref{eq_fluidity}).

As previously stated, steady-state solutions consist of three branches:\\
\begin{itemize}
\item[] branch ``0'' : $\dot \Gamma =0$, $a=0$, $\Sigma$ arbitrary. This corresponds
to a frozen, non-flowing system.
\item[] branch ``1'': $a < a_m$, $\dot\Gamma <\dot\gamma_m=a_m\sigma_m$, and $\Sigma$
a decreasing function of both $a$ and $\dot\Gamma$,
\item[] branch ``2'':
$a > a_m$, $\dot\Gamma >\dot\gamma_m$, and $\Sigma$
an increasing function of both $a$ and $\dot\Gamma$, from $\sigma_m$
up to $\infty$.
\end{itemize}
The three solutions (or branches) of fluidity  
are schematically plotted in the ($\Sigma$,$a$) plane on figure \ref{cur2}
(the $\Sigma(\dot\Gamma)$ plot has a similar aspect and is identical to the left curve in
figure \ref{cur}). We will use the notation $a_1(\Sigma)$ and $a_2(\Sigma)$ to designate branches ``1''
and ``2'' (the two roots of $\Sigma^2=f(a)/ah(a)$).

%%%%%%%%%%%%%%%%
\subsection{Imposed stress $\Sigma$}
%%%%%%%%%%%%%%%

The evolution of a homogeneous system at fixed $\Sigma=\sigma$ is
truly simple as the Maxwell model yields $\dot\Gamma=a\Sigma$
which can be injected in the equation describing the fluidity,
so that the evolution of the system 
is completely described by the sole differential equation: 
\begin{equation}
\frac{da}{dt} = -\frac{dV_\Sigma}{da}(a)
\label{eqpot}
\end{equation}
where the ``effective potential'' $V_{\Sigma}$ is :
\begin{equation}
 V_{\Sigma}(a)= \int_0^a da' [f(a')-a'h(a')\Sigma^2]
\label{eqpot2}
\end{equation}

The above equation describes the relaxation of $a$ in the potential
$V_{\Sigma}$. Depending on the value of the imposed $\Sigma$,
this potential has either one or three extrema:

\begin{figure}
\includegraphics[width=7.5cm]{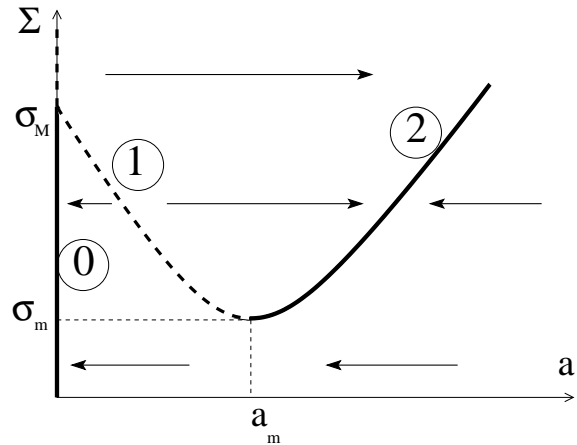}
\caption{Schematic description of the evolution of $a$ at fixed stress (horizontal arrows). 
Stable steady-state
branches are in bold,
unstable ones in dashed-lines.}
\label{cur2}
\end{figure}

\begin{figure}
\includegraphics[width=7.5cm]{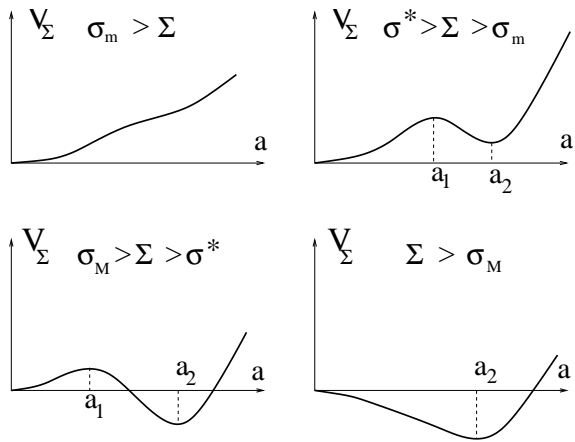}
\caption{
Effective potential $V_{\Sigma}(a)$ for the evolution of $a(t)$ in the homogeneous case.
Depending on the actual value of $\Sigma$, $V_{\Sigma}$ has one or two minima.}
\label{pot}
\end{figure}

\begin{enumerate}
\item[-] for $\Sigma < \s_m$, $a=0$ is the only minimum,
 
\item[-] for $\sigma_M>\Sigma > \s_m$, $V_{\Sigma}$ has two minima, $a=0$ and $a=a_2(\Sigma)$ and a
single maximum $a=a_1(\Sigma)$,
\item[-] for $\Sigma > \s_M$, $a=a_2(\Sigma)$ is the only minimum.
\end{enumerate}  

The linear stability immediately results (see Fig. \ref{cur2}), as the system falls in the potential
$V_{\Sigma}(a)$ along the instantaneous slope at point $a(t)$ (see Fig. \ref{pot}).

Branch ``0'' is stable for $\Sigma <\sigma_M$
so that the system evolves to this frozen 
situation whenever the stress is too weak $\Sigma <\sigma_m$.
At long time the system typically creeps but does not flow $\dot\Gamma(t)= a(t)\Sigma$
with $a(t)$ decreasing (typically algebraically in time)
to $0$. For example, with the explicit values of $f$ and $h$ proposed above,
one has $a(t) \sim 1/t$ so that $\Gamma(t) \sim \Sigma Log(t)$.

Branch ``1'' is always unstable, so that for $\sigma_M>\Sigma >\sigma_m$ the system
fluidity evolves either to $0$ (again leading to creep not steady flow)
if the initial value of $a$ is smaller than $a_1(\Sigma)$,
or towards $a_2(\Sigma)$ in the opposite case. Note that such a behaviour 
is rather similar to the recently
reported ``viscosity bifurcation'' \cite{cou3}. This similarity remains if the constraint of homogeneity
is lifted as will be discussed in section IV.
The stable branches ``0'' and ``2'' correspond to stable respectively non-flowing and flowing 
states of the system under $\Sigma$. 

For larger stresses $\Sigma>\sigma_M$ if the local flow curve is of type IV,
branch ``0'' is unstable and the fluidity always evolve towards $a_2(\Sigma)$.

Eventually we note that for $\sigma_M>\Sigma >\sigma_m$,
the addition of noise to the equation describing the evolution of 
$a$ could lead the system to select either branch ``0'' or `branch ``2'',
depending on which one corresponds to the lowest value of $V_{\Sigma}$
(the noise could allow the system to jump to the more favourable well, although possibly
after very long times).
Comparing these two values leads to introduce a value $\sigma^*$ that will prove
important in the following :
for $\sigma_m<\Sigma <\sigma^*$, $0=V_{\Sigma}(0) <V_{\Sigma}(a_2)$ so that 
branch ``0'' is the absolute minimum. For $\Sigma >\sigma^*$ branch ``2''
is the minimum. Formally, $\sigma^*$ is given 
by $V_{\sigma^*}(a_2(\sigma^*))=0$ (see Figure \ref{pot}).

%%%%%%%%%%%%%%%
\subsection{Imposed shear rate $\dot \Gamma$}
%%%%%%%%%%%%%%%

Although the steady-state solutions are obviously the same as in the previous case,
the dynamics is more complex when $\dot \Gamma$ is fixed,
as it is governed by the coupled equations for the evolution in time of
$\Sigma(t)$ and $a(t)$.

The stability of the steady-state solutions can however be examined
by linearisation around these solutions. 
The two stable branches $(0)$ and $(2)$ on figure \ref{cur2} described in the case of an 
imposed stress remain stable.
The stability of the decreasing branch $(1)$ at fixed shear rate and with homogeneity 
imposed depends on the choice of the functions $f$ and $h$ in equation (\ref{eq_fluidity_d2}) 
(recall that as mentioned in section \ref{sub2}  we have formally suppressed 
the mechanical inertial mode of destabilisation). 
For general $f$ and $h$, one can not preclude a destabilisation of 
the steady solution $(1)$ to a limit cycle \cite{are}. 
However, this is not the case with the expressions of section \ref{sub1} $f(a) = r a^2 + a^4 $ 
and $ h(a)=r'a+a^2$, for which the branch $(1)$ is always stable (again when homogeneity is enforced).

%%%%%%%%%%%%%%%%%%%%%%%%%%%%%%%%%%%%%%%%%%%%%%%%%% 
\section{Heterogeneous flow at imposed stress $\Sigma$.}
\label{stress_H}
%%%%%%%%%%%%%%%%%%%%%%%%%%%%%%%%%%%%%%%%%%%%%%%%%%

The situation at imposed fixed stress can be apprehended rather simply
starting from the analysis of the homogeneous case mentioned above
with its two stable branches ``0'' and ``2''. 
Heterogeneities are generated by the 
walls which induce wall layers of different fluidities,
with a major effect on the selection of the steady-state flow branch.

%%%%%%%%%%%%%%%%%%%%%
\subsection{Imposed fluidity at the wall}
%%%%%%%%%%%%%%%%%%%%%%

We start by considering the case where the walls impose
the local value of the fluidity
$a|_{z=0,H}=a_w$.
The evolution of the system is described by that of
$a(z,t)$ from which the shear rate can easily be deduced 
$\dot\gamma(z,t)=a(z,t) \Sigma$. Formally,
\begin{equation}
\frac{da}{dt}= - \frac{\delta F_{\Sigma}}{\delta a(z)}
\end{equation}
which is a rather classical equation describing the local relaxation
of a ``free energy'' $F_{\Sigma}({a(z)})$. Here this functional is:
\begin{equation}
F_{\Sigma}({a(z)})= \int_0^H [V_{\Sigma}(a) + \frac{1}{2}D(\partial_z a)^2] dz
\end{equation}
where it is implicitly understood that $a$ equals $a_w$ on the boundaries.

The evolution can be understood as a minimisation of $F_{\Sigma}$,
as $d_tF_{\Sigma}= -\int_0^H dz (\partial_ta)^2 \,<0$.
Schematically, minima correspond to adopting throughout   
the sample a value of $a$ minimising $V_{\Sigma}$ (i.e. $0$ or $a_2(\Sigma)$),
and then connecting at the boundaries
to the imposed value $a_w$ with a minimal penalty obtained as a compromise between
the $V_{\Sigma}$ term and the square gradient term (a more
formal description of these steady-state solutions is given in the Appendix).
This leads to steady-state branches in the ($\Sigma, \dot \Gamma$) plane that deviate
from the branches ``0'' and ``2'' only by the contribution of these 
wall layers, that shifts the average shear rate 
by a small term proportional to $\ell/H$, where $\ell \sim (Da_w/f(a_w))^{1/2} $ 
is the thickness of these layers.

\begin{figure}
\includegraphics[width=7.5cm]{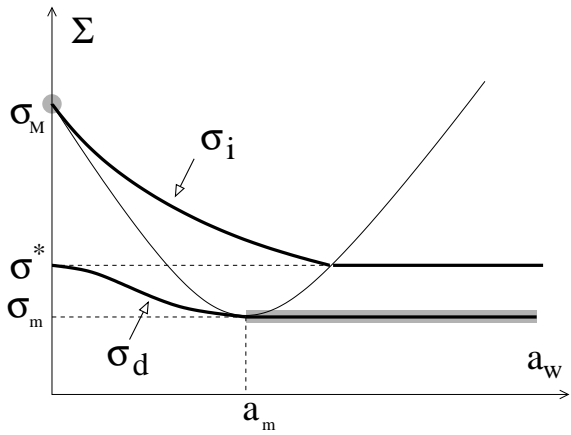}
\caption{ Thresholds values $\sigma_i$ and $\sigma_d$
of the stress as a function of the fluidity at the walls $a_w$.
Above $\sigma_i$
the frozen branch is unstable, below $\sigma_d$ 
the fluidised branch is unstable. When the imposed macroscopic stress
$\Sigma$ is between $\sigma_d$ and $\sigma_i$, both branches are stable.
Which one is selected depends then on the initial conditions, i.e. on the system's history.
Areas in grey correspond to bulk destabilisation (``spinodal''), whereas 
a simple thick line denotes destabilisation from the wall (``heterogenous nucleation'')}
\label{cur3}
\end{figure}

The stability of these branches is
now limited by the fact that the walls at $a_w$ can act as nucleating
sites for the other phase/branch.
%\noindent
For example the ``frozen'' branch, where fluidity decreases from $a_w$ on the wall to almost
zero in the middle, ceases to be stable if $\Sigma$ is increased 
so that  it becomes interesting (in terms of minimising $F_\Sigma$) to increase $a$ beyond $a_w$
at the expense of a section of the $a\simeq 0$  central part.
The threshold roughly corresponds to $V_{\Sigma}(a_w)=V_{\Sigma}(0)$
which defines a maximal value of the stress $\sigma_i(a_w)$ 
for stability on the pasty branch. Beyond  this value, the $a(z)$ solution
``slips'' into the potential well of branch ``2''.
%\noindent
Similarly, the fluid branch ceases to be stable upon decrease of the stress $\Sigma$ 
when the latter reaches the value $\sigma_d(a_w)$ defined by
$V_{\sigma_d}(a_w)=V_{\sigma_d}(a_2(\sigma_d))$.
It is easy to check that $\sigma_m \le \sigma_d \le \sigma^*$,
whereas $\sigma^* \le \sigma_i \le \sigma_M$.
Typical curves for $\sigma_i(a_w)$ and $\sigma_d(a_w)$
are plotted in figure \ref{cur3}.

A second process of destabilisation exists. If the wall fluidity has extreme values (either low or high) walls may no longer be involved in the destabilisation process which is then ``spinodal" (grey areas in figure 6).

\begin{figure}
\includegraphics[width=7.5cm]{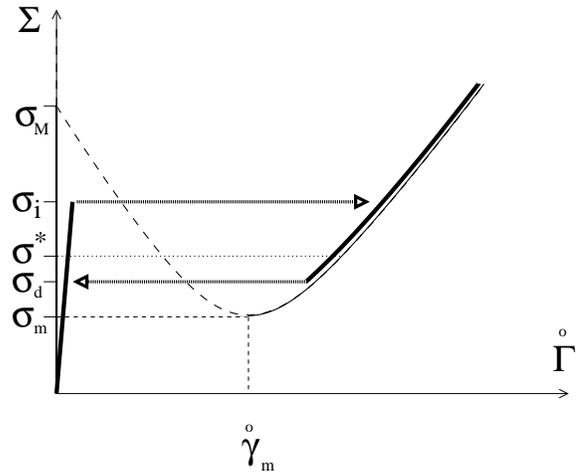}
\caption{ Macroscopic flow curve at imposed stress $\Sigma$ (thick lines).
The thin lines (dashed and continuous ) correspond to the local flow curve of Figure 2.
The value of the fluidity $a_w$ 
at the walls fixes the location of the limit of stability $\sigma_i$ and $\sigma_d$
of the frozen and fluid branches (see Figure 6). When the stress is varied monotonously, the shear rate
suddenly jumps at these values (thick arrows).
}
\label{cur4}
\end{figure}

The conclusion for the macroscopic steady-flow $\Sigma(\dot\Gamma)$
diagram is thus simple.
At fixed $\Sigma$ there are two stable flow curves:\\
- for $\Sigma <\sigma_i(a_w)$ a branch close to the axis corresponding
to a frozen bulk with two layers of finite fluidity close to the walls
(as schematised e.g. in Figure 2).
This branch is indistinguishable from the vertical axis
for large values of the thickness $H$.\\
- for $\Sigma >\sigma_d(a_w)$, a fluidised branch very close to ``2''
(identical in the limit $H \rightarrow \infty$),
with a flowing bulk, and slightly less fluid wall layers. 

In the intermediate range $\sigma_d < \Sigma <\sigma_i$,
both branches are stable.
The branch actually selected depends on the system history and preparation (initial conditions). 
This is very compatible with recent observations of a ``bifurcation'' of the asymptotic
viscosity at a given value of the applied stress for a given preparation \cite{cou3}.

If one starts with macroscopic bands of fluidised and frozen material, 
the fronts between these zones will move so that one typically ends on the 
fluid branch for $\Sigma > \s^*$,  and on the frozen branch for $\Sigma < \s^*$.

If the stress $\Sigma$ is varied slowly in a systematic way (i.e. increased or decreased),
one has the simple hysteretic behaviour described by the thick arrows in Figure \ref{cur4}.

The departure from the homogeneous case described in the previous section
is due to the walls, which introduce
the parameters $H$ and $a_w$. Although the former has a moderate role in shifting
by a small amount the branches in the diagram (by a term proportional to $\ell/H$), 
the latter controls the position
of the instability thresholds,
and thus the amplitude $\sigma_i -\sigma_d$ of the hysteresis loop
observed macroscopically
(actually looking into details, these thresholds stresses
 depend very slightly on $H$ but this effect is almost invisible numerically. See Appendix).

\begin{figure}
\includegraphics[width=8.5cm]{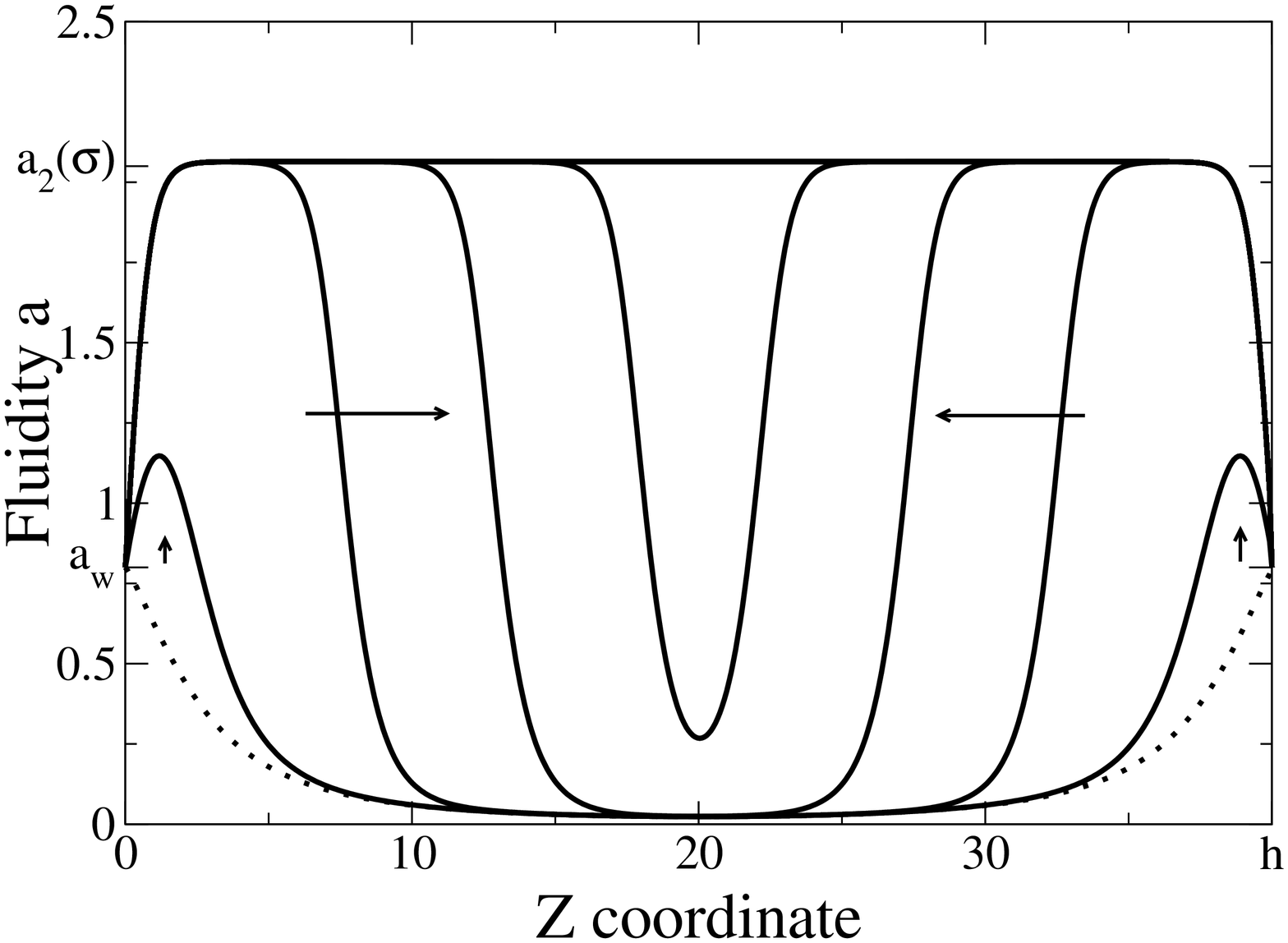}
\caption{Destabilisation of the 'frozen bulk' by heterogeneous nucleation. 
The initial state is plotted with dotted lines. As the stress is increased so that $\Sigma \gtrsim \s_i$, 
two fronts come from the walls, as shown by the arrows  before they merge in the centre. 
The steady state is the fluidised bulk.
 In this simulation $f(a)= a^2 +a^4, h(a)=0.05*a + a^2, a_w=0.8$, $\Sigma=1.45$}.
\label{frontmerging}
\end{figure}

The above picture (Figure \ref{cur4}) is borne out by numerical simulations of 
the equations for various initial conditions,
and also submitting the system to ramps
of increasing and decreasing stress. 
These simulations also allow to follow the destabilisation process.
Starting from a frozen solution, the bulk is destabilised when $\Sigma=\s_i$.
This destabilisation process is initiated at the wall, where the fluidity is fixed and non vanishing
(acting therefore as a destabilising nucleus) : two fronts of large fluidity are first created at each wall,
then propagate in the material to merge eventually at the middle of the Couette cell to give the
homogeneous $a_2(\Sigma)$ bulk solution (Figure \ref{frontmerging}).

On the other hand, starting from the fluidised branch, when the 
stress is decreased this highly sheared solution is destabilised for $\Sigma =\s_d$.
The destabilisation process depends on $a_w$. 
If $a_w < a_m$, then $\s_d > \sigma_m$ and the destabilisation occurs by fronts merging, as 
previously described, 
whereas in the reverse case  $\s_d=\s_m $, and the bulk is destabilised as a whole, as a plateau 
decreasing in time towards zero.

Note that the problem is formally very similar to that of wetting transitions in thermodynamics: 
it reduces to the minimisation of a free energy functional that is the sum of 
squared gradient term $(\partial_z a)^2$ and a potential with two wells $V_{\sigma}(a)$, 
with boundary conditions on the walls that can favour one or the other of the 'phases'.

In this interpretation,  the sections $\sigma^* \rightarrow \sigma_i$ on the pasty branch, 
and  $\sigma_d \rightarrow \sigma*$ on the flowing one would be called metastable.
The absence of noise in the equations used here makes them absolutely stable. 
Adding some noise in the bulk equation for $a$ would reduce the metastable branch, 
bringing the system (possibly after prohibitively long waiting times)
to jump from the pasty to the fluid branch at $\sigma*$.

Note that it may be experimentally difficult to detect the hysteresis loop if 
its amplitude is small (either because $\sigma_i -\sigma_d$ is small
or because of the presence of noise), which would
make this macroscopic flow curve very similar to a textbook description
of a fluid with a yield stress $\s^*$, with possibly a rather abrupt change of the steady-state shear rate
at this value.

%%%%%%%%%%%%%%%%%%%%%%%%%
\subsection{No flux at the wall : $\partial_z(a)=0$}
%%%%%%%%%%%%%%%%%%%%%%%%%%

If the boundary conditions are such that $\partial_za=0$ on the walls,
then the behaviour of the system is exactly that 
obtained in the analysis of the homogeneous model:
the system tends to relax any inhomogeneity,
and its stable steady-state solutions 
are those of the homogeneous model:
a pasty branch ``0'' for $\Sigma <\sigma_M$ and a 
fluid branch ``2'' for $\Sigma >\sigma_m$.
Both branches are stable solutions for $\sigma_m<\Sigma<\sigma_M$.
A hysteretic loop results of amplitude larger than in the previous case.

So the picture is still quite similar to Figure 7 above with the following
differences: here $\sigma_i=\sigma_M$ and $\sigma_d=\sigma_m$,
and the branches follow strictly the local flow curve, with no slight deviation
due to wall layers.

Again we have checked the validity of this picture numerically.

%%%%%%%%%%%%%%%%%%%%%%%%%%%%%%%%%%%%%%%%%%%%%%%%%%%
\section{Heterogeneities at controlled shear rate $\dot \Gamma$}
\label{shearrate}
%%%%%%%%%%%%%%%%%%%%%%%%%%%%%%%%%%%%%%%%%%%%%%%%%%%

We now turn to the case where the global shear rate is controlled, \ie the velocity $V$ of the top
wall is fixed, so that the average shear rate $\dot \Gamma$ is imposed.
The dynamics of the system is now more complex and governed by two coupled integro-differential
equation, that can be obtained from (\ref{eq_fluidity_d2}) :

\begin{equation}
\left\{ 
\begin{array}{ll}
\partial_t \Sigma = -\left< a\right>(t)\Sigma(t)  + \dot \Gamma \\
\partial_t a = - f(a) + ah (a) \Sigma^2(t)  + h(a)\Sigma \partial_t\Sigma + D \partial^2_{zz}a\\
\end{array} 
\right.
\label{eq_const-g}
\end{equation}
where $\left< a\right>$ is the spatial average of $a(z,t)$,
\begin{equation}
\left< a\right>(t)=\frac{1}{H} \int_0^H dz \, a(z,t) \, ,
\label{gamaverage}
\end{equation}
and where the boundary conditions are either
$a|_{z=0,H}=a_w $, or $\partial_z a|_{z=0,H}=0$.
The local shear-rate is given by $\dot\gamma (z,t)=\dot\Gamma +(a(z,t)-\left< a\right>(t))\Sigma$.

This set of dynamic equations leads to a much richer set of situations, 
as the dynamics of $a$ does not relax a free energy
(formally, $\partial_t a= - \delta F_{\Sigma(t)}/\delta a(z) +\frac{1}{2}h(a)\partial_t \Sigma^2$).
As a result, for some values of $\dot\Gamma$, the system can have 
many steady-states corresponding to different values of $\Sigma$,
and for some others none (the system then tends to a limit cycle corresponding to a periodic
solution for $a(z,t)$ and for $\Sigma(t)$  that yields a macroscopic stick-slip behaviour). 

To explore this ensemble of scenarios it is necessary to solve numerically these equations.
Our numerical exploration leads to a complex picture.
Starting with the case of a fixed fluidity at the walls, 
we first describe the main features observed, before pointing out 
more complex features, including the metastability of various patterns of shear bands
and the occurrence
 of localised oscillations. At this stage we have no analytical understanding of some of these 
features, and of their dependence on the parameters. The behaviours reported here result of numerical simulations with $f$ and $h$ fixed to  $f(a)= r\,a^2+a^4$, and $h(a)=r' \,a + a^2 $.
At the end of this section we turn to the case of zero fluidity gradient
at the walls to underline a few differences with the previous case.

%%%%%%%%%%%%%%%%%%
\subsection{Imposed fluidity at the wall.}
%%%%%%%%%%%%%%%%%%

\begin{figure}
\includegraphics[width=7.5cm]{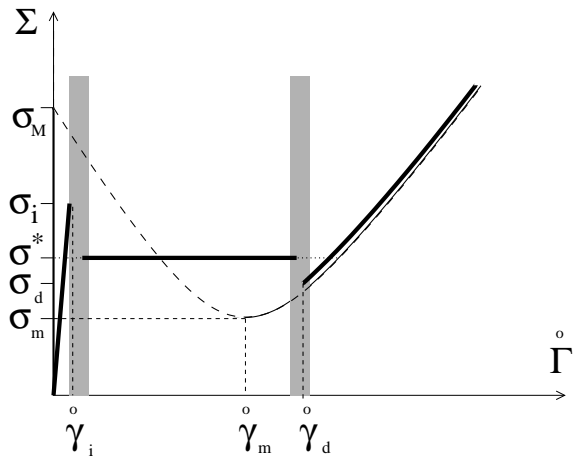}
\caption{Schematic macroscopic flow curve at imposed shear rate $\dot\Gamma$ (thick lines).
Again the thin lines (continuous and dashed) denote the local flow curve.
The value of the fluidity at the walls fixes the 
location of the limit shear rates $\dot\gamma_i$ and $\dot\gamma_d$
of the frozen and fluid homogeneous branches.
Between these values the system tends to segregate into shear bands.
At the transition between the two (grey areas), stable or oscillating
thin layer structures are observed in the vicinity of the walls.
}
\label{cur5}
\end{figure}

We start by analysing the case where the fluidity 
at the walls is fixed to a value $a_w$.
Depending on $\dot \Gamma$ and $a_w$ different behaviours are observed.
The main features are the following:\\
- The frozen $0<\Sigma<\sigma_i$ and fluidised $\sigma_d <\Sigma$  branches found at
fixed stress (section \ref{stress_H}), are also found to be stable at imposed $\dot \Gamma$.\\
- Between two values $\dot\gamma_i$ and $\dot\gamma_d$ defined in terms of the
threshold stresses on these branches (see fig. \ref{cur5}), the most classical long time behaviour 
(which we take as steady-states), are solutions corresponding to the coexistence
of ``bands'' of frozen ($a\simeq 0$) and of fluidised phase ($a\simeq a_2(\sigma^*)$)
coexisting at a stress $\Sigma \simeq \sigma^*$.\\
- In a narrow vicinity of the transition from the quasi-homogeneous
branches to this situation of coexisting ``shear bands'' (i.e. $\dot\Gamma$
slightly larger than $\dot\gamma_i$ or slightly smaller  than
$\dot\gamma_d$), various phenomena can occur in the vicinity of the wall:
nucleation of thin ``layers'' of fluidity different from the bulk, or of  ``layers'' of periodically oscillating fluidity.

These behaviours have different signatures as shown on the macroscopic flow curve sketched on figure \ref{cur5}.

%%%
\subsubsection{Frozen and fluidised regimes}
%%%

First, for small and large imposed shear rate, $\dot \Gamma$ one recovers the solutions
obtained at imposed stress : except for small boundary effects,
they correspond to the frozen branch "0'', and to the
 fluidised branch ``2'' of the local flow curve.

Along the fluidised branch, the $\Sigma(\Gamma)$ relationship 
almost exactly follows the intrinsic
flow curve obtained in the homogeneous case.
The frozen branch actually displays an almost vertical
 section going to zero in the limit of
vanishing shear rate $\dot \Gamma \rightarrow 0$ due to the
presence of a small wall layer of finite fluidity imposed by the wall, $a_w$,
and of thickness $\ell$.

These domains are limited by the global stresses $\s_i(a_w)$ and $\s_d(a_w)$, 
as discussed in section \ref{stress_H}. 
The corresponding shear rates $\dot\gamma_i$ and $\dot\gamma_d$
can be calculated numerically with the method described in the Appendix.

%%%%%
\subsubsection{Shear-banding regime.} 
%%%%%

\begin{figure}
\subfigure[]{\includegraphics[width=8.5cm]{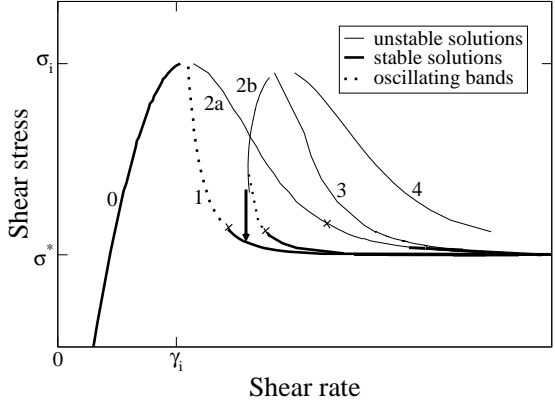}}
\subfigure[]{\includegraphics[width=8.5cm]{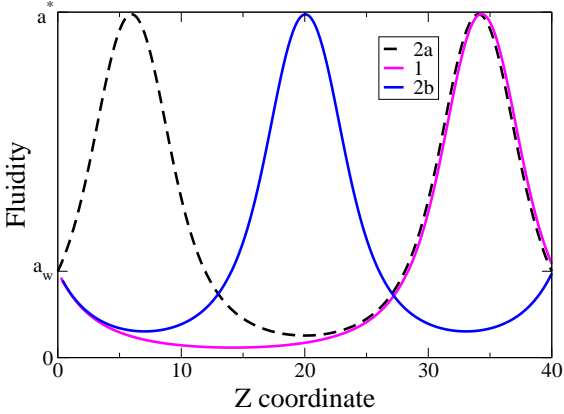}}
\caption{ (a)Calculated steady state flow curves : 
the unstable stationary solutions are plotted with thin lines, 
the stable solutions with thick lines, the oscillations with dotted lines. 
The numbers on the branches represent the number of 
fronts between the fluidised and frozen bands. $r=1, r'=0.01, H=40, a_w=0.25$. 
b) Fluidity profiles corresponding to the points marked with crosses 
on figure (a) for the branches 2a,2b, and 3. }
\label{zoom}
\end{figure}
A shear-banding domain connects the two previous frozen and fluidised regions. 
It corresponds to a "coexistence" in the shear cell of the frozen and of the fluidised states. 
Its signature on the macroscopic flow curve is a  {\it quasi-plateau} at a stress $\Sigma \sim  \s^*$.  
All over the banding domain (for a given $a_w$), the stress is found to remain in the vicinity 
of the value $\s^*$ defined in section \ref{homogeneous} by 
$V_{\s^*}(a_2(\s^*))=V_{\s^*}(0)$. This approximately fixes the value of the fluidity 
in the fluidised region to $a^*=a_2(\s^*)$, with $a_2(\s)$ being the corresponding branch 
obtained in the homogeneous case (see section \ref{homogeneous}). The
shear rate in the fluidised region is approximately given accordingly  as $\dgam^*=a^*\s^*$.
The total width of the frozen and fluidised regions are determined by the global constraint 
on the total shear rate. The width of the fluidised region $h_f$  
is typically given by $h_f \simeq H \dot \Gamma /\dgam^*$, with $\dgam^*$ defined here above
(small boundary effects neglected).

An interesting question concerns the number of fluidised bands : how many fronts between 
a frozen and a fluidised bands can be stabilised for given $H$, $a_w$, and $\dot \Gamma$ ? 
\noindent
In order to answer this question, a 'zoom' of the {\it quasi plateau}  is plotted on figure \ref{zoom} (a) for given $H$ and $a_w$. The stable {\it quasi plateau} actually  consists of parts of the branches representing the steady states solutions with several fronts (on the figure these stable parts of the branches are plotted with thick lines).
The steady state multi front branches were calculated numerically with the method detailed in the appendix. The fluidity profiles corresponding to the branches with one and two fronts are represented on figure \ref{zoom} (b). Their  stability was then examined with the numerical simulation described in \ref{sumup}.

\begin{figure}
\includegraphics[width=8.5cm]{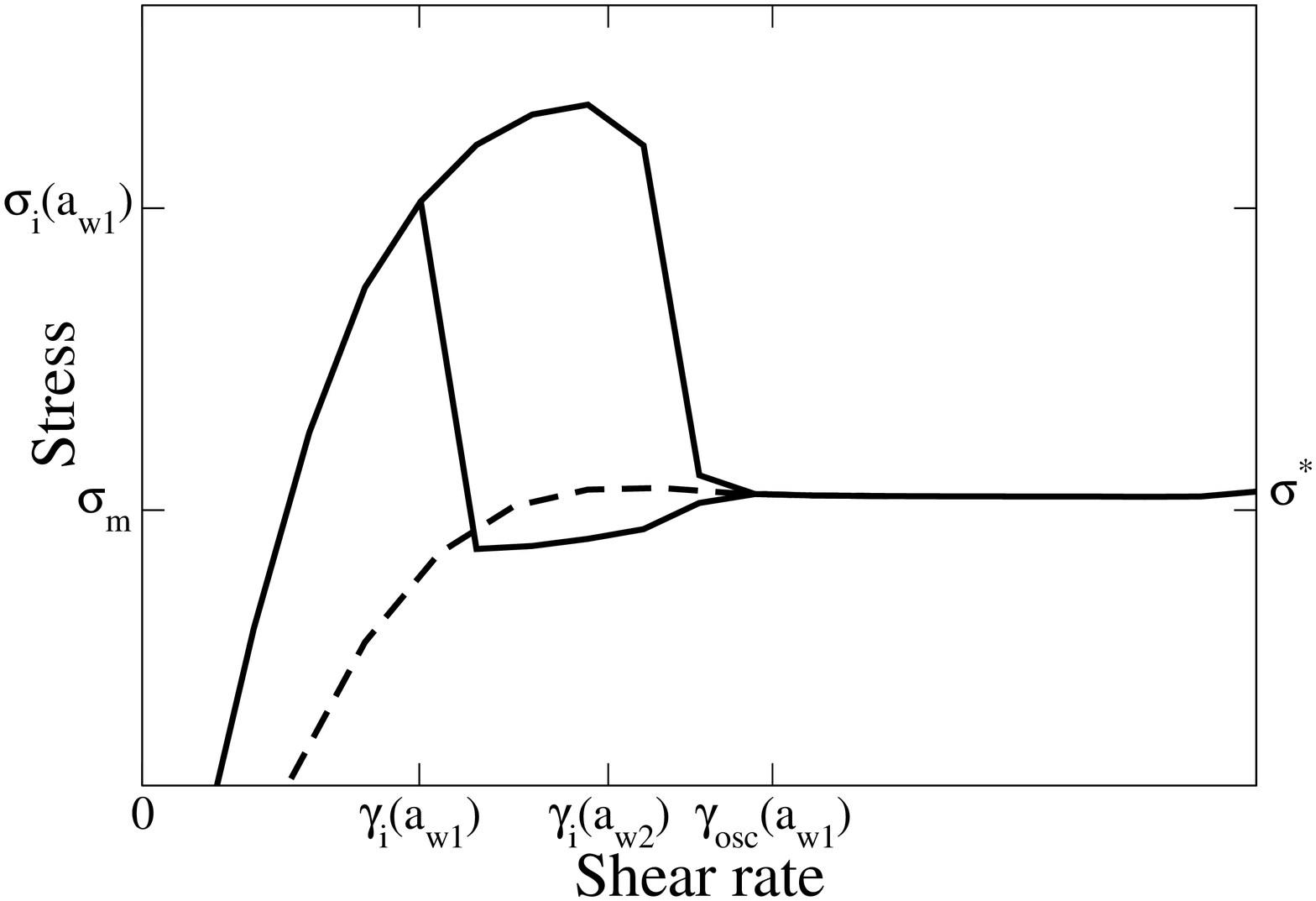}
\caption{Behaviour at low shear rate for a fixed $H=100$, $r=1,r'=0.01$. 
The solid line corresponds to $a_{w1}=0.3$ and the dashed line to $a_{w2}=1$.
In the case of  $a_{w1}$, oscillations are observed, 
and the envelope of the oscillations amplitude is represented}
\label{stabaw}
\end{figure}

\noindent
No analytical characterisation was found to the limits of stability of the multi front branches. 
Beyond these limits, either oscillations are observed, or other stable branches are reached.
Their complex dependence on $a_w$ and $\Sigma$, is shown on figure \ref{stabaw}.
Note that, with the parameters taken to plot figure \ref{zoom} (a), 
for a small range of $\dot {\Gamma}$ (just left of the middle cross
in Figure 10a), 
two solutions are possible (depending on the history of the system) : 
stable oscillations of a central fluidised layer (branch 2b), 
and a steady solution with a single fluid layer at the wall (branch 1). 
In the case of oscillations, when $\dot {\Gamma}$ is 
further decreased to the limit of stability of branch 2b, the system evolves 
toward the pattern of branch 1 as indicated by the arrow in Figure 10a.

%%%%%%%
\subsubsection{Stick slip behaviour} 
\label{stsl}
%%%%%%%
When no stable branch is reached, a time dependent oscillating behaviour is 
evidenced for stress and fluidity.
As seen on figure \ref{stabaw}, this occurs for small wall fluidity and imposed shear rate. 
This behaviour is very reminiscent of stick slip, a phenomenon commonly encountered in solid friction
\cite{bom,are}. 
Here it originates in an oscillating behaviour of the fluidity 
between a "frozen" state ("stick" phase) and a partially fluidised state ("slip" phase). 

\begin{figure}
\subfigure[]{\includegraphics[width=8.5cm]{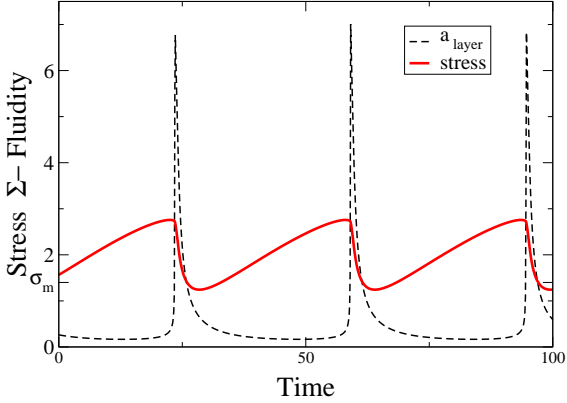}}
\subfigure[]{\includegraphics[width=8.5cm]{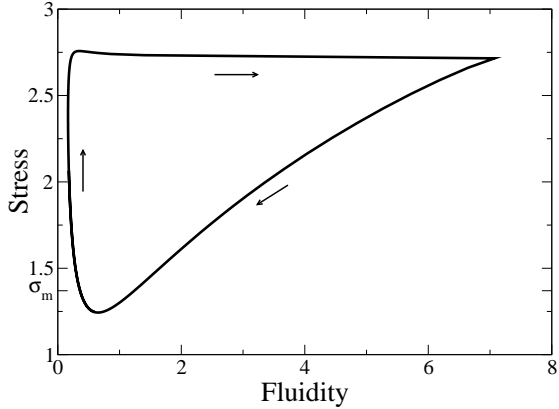}}
\caption{ The amplitude of the oscillations of the global stress and of the fluidity of the oscillating layer are represented as a function of time (a), and in the plane $(a,\Sigma)$ (b), in the case $a_w=0.3$, $r=1, r'=0.01, \dot{\Gamma}=0.12$. Its width hardly varies.}
\label{omeca}
\end{figure}

The fluidisation however occurs only within 
one layer, the rest of the system being frozen. This spatially 
localised oscillating behaviour resembles a periodic fracture-healing process,
as observed {\it e.g.} by Pignon \etal \cite{pig}. The fluidised layer is localised either close 
to the wall, or in the middle of the shear cell,
its fluidity oscillates between two extremal values, and its thickness is approximately fixed. 
The time evolution of $\Sigma$ and the maximum fluidity of the layer are plotted 
on figure \ref{omeca}, showing characteristic features of a "stick-slip" behaviour. 

The system rapidly relaxes to the frozen state,  \ie a 
very small fluidity, while the stress increases until a limit where the fluidity abruptly increases, \ie the 
fluid is suddenly sheared. This sudden increase is due to the penetration of the maximum fluidity in the 
well potential of $a_2(\s)$, that is enlarged as $\s$ increases. To maintain a fixed global shear rate, 
$\s$ relaxes quickly and the cycle starts again.

%%%%%%%%%%%%%%%%%%%%%%
\subsection{No flux boundary condition at the wall : $\partial_z a |_{z=0,H}=0$}
%%%%%%%%%%%%%%%%%%%%%%

In the case of a no flux boundary condition and for intermediate values of $\dot \Gamma$ 
($\dgam_i < \dot \Gamma < \dgam_d$), two very different locally stable 
sets of solutions of (11,12) are possible :
one is very similar to the solutions obtained above for fixed  
fluidity at the walls and exhibits shear banding and "stick-slip" behaviour,
the other one is the homogeneous solution at fixed shear rate obtained in section \ref{homogeneous}. 
This peculiarity is associated with the specific form of the boundary condition,
which is compatible with a homogeneous fluidity in the cell. Moreover, as discussed
in section III, the decreasing branch is linearly stable at fixed shear rate (remember that
we have suppressed inertial terms), and no
destabilising mechanism is induced here by the boundaries (in contrast with the fixed wall fluidity case).
Formally, the system is thus attracted either by the homogeneous solution or by the inhomogeneous one,
depending on initial conditions 
(an initial inhomogeneous state often leading to a banded inhomogeneous solution).

The macroscopic flow curve for (11,12) with the present boundary conditions consists of: 
\begin{itemize}
\item[-] the two increasing branches of the local curve ( the branches are parts of the local curve as the solutions are homogeneous). Those branches are also stable in the case of imposed stress.
\item[-] the decreasing branch of the local curve. This is of course an artefact due to 
our neglecting inertial terms, which suppresses the``mechanical instability'' 
that always destabilises this branch. 
\item[-] the {\it quasi plateau}, with $\Sigma \sim \s^*$. As in the case of an imposed fluidity at the wall, 
it is constituted of parts of multi front branches, that become unstable at different values of the shear rate.  
\item[-] an oscillating domain, for small shear rates, when the single front branch looses its stability.
\end{itemize}
Ruling out the unphysical decreasing branch, the picture is thus quite similar to Figure 9 (without the
effective wall slip in branch 1 due to the finite wall fluidity there).  

%%%%%%%%%%%%%%%%%%%%
\subsection{Comment on oscillations}
Oscillating response under steady driving has recently been numerically observed  in a
model for shear-thickening materials \cite{hea02}, which in contrast to the present
one described the homogeneous behaviour of a system (no spatial heterogeneity
in the formalism), and the oscillations were shear-rate oscillations at fixed stress.
Formally there are some similarities
as in both cases a continuous (and thus infinite) set of variables
evolves through coupled differential equations
with non-monotonic
functions involved, and driven by a global integral constraint.
However, in contrast to the situation of \cite{hea02} where all the variables
were involved in the oscillation, here only a limited set
of variables do (corresponding to a spatially localised oscillating layer).

In principle a finite number of variables is enough to get an oscillating rheological
response (limit cycles) under steady driving \cite{cat02}, so it is no surprise
oscillations show up here given the complexity of the set
equations (\ref{eq_const-g},\ref{gamaverage}) and the number of variables involved. One may even
expect more complex dynamical behaviour such as  some form of chaotic
response \cite{cat02}, but we found none within the set of parameters numerically explored.
Clearly, a truly three dimensional version of this model, with the corresponding non-linear
convective terms, should increase the likeliness of turbulent/chaotic behaviour (even in
the absence of inertia \cite{gro98}).

%%%%%%%%%%%%%%%%%%%%%%%%%%%%%%%%%%%%%%%%%%%%%%%%%
\section{Discussion}
\label{discussion}
%%%%%%%%%%%%%%%%%%%%%%%%%%%%%%%%%%%%%%%%%%%%%%%%%

In the present paper we used an extension of a model previously introduced
to account for the rheology and aging in pasty systems, in which the steady state local flow curve is non-monotonous so that heterogeneous flows can be generated at steady state. The key ingredient of the model is a coupling between
the mechanical properties, namely  shear stress or shear rate, and local relaxation within the fluid, characterised by a "fluidity". 
Although analytically very simple, the model yields a rather rich macroscopic phenomenology.

\begin{itemize}
\item Under controlled stress $\Sigma$, the steady state behaviour is schematised on figure \ref{cur4}. 
The fluid undergoes an hysteretic phenomenon if the stress is increased or decreased progressively. 
The effective yield stress is 
shown to depend on the fluid history and on the wall conditions. Besides, no banding is observed and 
there is a discontinuity in 
shear rate $\dot{\Gamma}$ which occurs at specific values of the stress.

In this case, we have shown that the dynamics reduces to the minimisation of a free energy function. 
This is generalisable to any function $f$, $g$ describing aging and fluidisation, and to more complex 
diffusion terms $D(a)$.
Note that the dynamics was not built as a free energy relaxation but naturally appears this way 
(see related discussions in \cite{Harrowel}). This gives some generality to the qualitative results, 
beyond the specific model studied here.

As no noise term was included, the equivalent of homogeneous nucleation in thermodynamics problem 
was suppressed and boundary conditions play a crucial role in destabilisation processes. 
Hence, the values of the thresholds depend on the boundary conditions. 
However, the qualitative behaviour (hysteretic loop) is rather insensitive to the specific 
type of boundary condition chosen.

\item At fixed global shear rate $\dot{\Gamma}$, the macroscopic behaviour is schematised 
on figure \ref{cur5} :
 banding solutions show up, separating a "frozen" region
from a fluidised region. The
shear rate within the fluidised band is hardly dependent on the imposed global shear rate, 
and fixed only by characteristics of the material. 
Within the banding domain, the thickness of the
fluidised region increases (nearly) linearly with the global shear rate $\dot{\Gamma}$ 
to eventually fill the full shear cell.

At small (total) shear rate, an oscillating behaviour of the stress $\Sigma$ is observed, 
with features very reminiscent of a classical "stick-slip" behaviour. The system
is globally in a frozen state, and fluidises periodically within a thin layer. Such a 
behaviour compares well with the fracturing/ healing process observed by 
Chen \etal \cite{che} and Pignon \etal \cite{pig}. These localized oscillations are to be
contrasted with homogeneous oscillations obtained in other models for different
systems (e.g. \cite{hea02}).

Our observations result from numerical simulations and the influence of the specificity 
of the modelisation of the fluidisation and aging remains to be understood.
Again, the different thresholds depend on the boundary conditions, but the qualitative 
behaviour described are robust to changes in the types of boundary conditions. Extension
of the present model to a fully three-dimensional description
could however yield a more complex dynamical behaviour.
\end{itemize}

In conclusion, we suggest that analysing yield stress fluids close to the yield stress at both 
constant stress and constant shear rate could provide interesting information. It could also 
prove informative to check the sensitivity of the results to the nature of the surfaces bounding the fluid.

%%%%%%%%%%%%%%%%%%%%%%%%%%%%%%%%%%%%%%%%%%%%%%%
\appendix

\section{Appendix : Stationary solutions of equation (\ref{eq_fluidity_d2})}
In this appendix, we detail the steady state solutions of Eq. (\ref{eq_fluidity_d2}) recalled here  :

\begin{equation}
\left\{ 
\begin{array}{ll}
a(z) \Sigma = \dot \gamma(z)\\
 -f(a(z)) + a(z)h(a(z))\Sigma^2 + D \partial^2_{zz}a(z)=0\\
a|_{z=0,H}=a_w \,\,\, {\rm or}\,\,\, \partial_z a|_{z=0,H}=0  
\end{array} 
\right.
\end{equation}
For clarity, the potential  $W_{\Sigma}(a)$ is introduced as :

$ W_{\Sigma}(a)= -\int_0^a da' [f(a')-a'h(a')\Sigma^2]$.

Note that $W_{\Sigma}=- V_{\Sigma}$ as defined in Eq. (\ref{eqpot2}).
Hence for a given $\Sigma$, we look for a solution $a(z)$ such that 
\begin{equation}
\left\{ 
\begin{array}{ll}
-\partial_a W_{\Sigma}(a)= D \partial^2_{zz}a(z)\\
a|_{z=0,H}=a_w \,\,\, {\rm or}\,\,\, \partial_z a|_{z=0,H}=0  
\end{array} 
\right.
\label{eqpot3}
\end{equation}

the local shear rate is then immediately deduced : $\dot \gamma(z)= a(z) \Sigma$.

Equation (\ref{eqpot3}) can be interpreted as the classical equation of motion 
of a particle of mass $D$ and position $a$ in the potential $W_{\Sigma}$ 
where $z$ plays the role of the time.
The energy $ E_m= \frac{1}{2}(\partial_z a)^2  +W_{\Sigma}(a)$ is conserved during the motion.
\begin{figure}
\subfigure[]{\includegraphics[width=8.5cm]{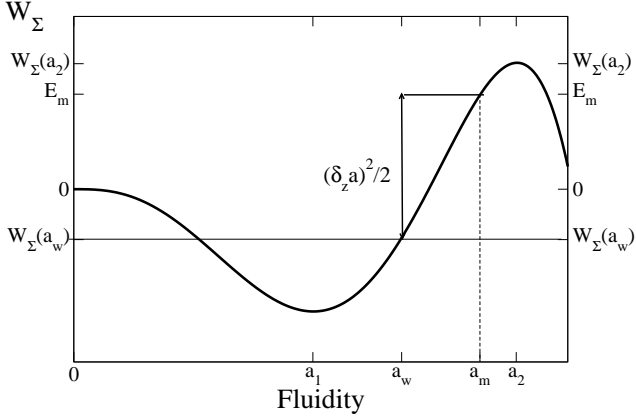}}
\subfigure[]{\includegraphics[width=8.5cm]{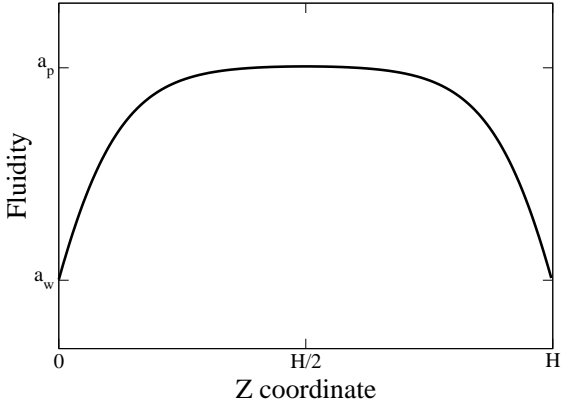}}
\caption{a) Potential $W_{\Sigma}$. $\Sigma=1.44, r=1,r'=0.01$. 
b) Fluidity profile for a fluidised bulk. $a_w=1, H=15$.}
\label{figpot}
\end{figure}

With this interpretation, the boundary conditions impose either initial and final positions 
($a_w$ imposed at the walls), or velocities ($\partial_z a=0$ imposed at the walls).

We choose to detail here the case of an imposed $a_w$ at the wall, and  the stress $\Sigma$ 
is chosen $\s_i > \Sigma > \s_d $ (see section \ref{stress_H}).

The potential $W_{\Sigma}$ is represented on figure \ref{figpot}. The extreme values 
of the `mechanical energy'  $E_m$ are $W_{\Sigma}(a_w)$ and $W_{\Sigma}(a_2)$. 
From the analysis of these solutions, it is possible to progress further in answering a few questions : 

\begin{itemize}
\item {\it What is the fluidity profile on the frozen and fluidised branches of figures \ref{cur4} 
and \ref{cur5}?}

For a given thickness $H$, $E_m$ is set so that $a|_{z=0,H}=a_w$ with $\partial_z a|_{z =0}> 0$. 
With the mechanical analogy, this amounts to fix the mechanical energy or initial velocity for 
the particle to leave position $a_w$, reach a maximum $a_m$ with  $\partial_z a|{a_m}=0$, 
and go back to its initial position within a time $H$.

The fluidity profile is represented on figure~\ref{figpot}.
Note that  $E_m=W_{\Sigma}(a_2)$ corresponds to a fluidised bulk, 
with an infinite thickness $H \sim \infty$.

With $\partial_z a|_{z =0}> 0$, a profile with a bulk of fluidity higher than $a_w$ is described. 

Symmetrically, a profile on the frozen branch, that is with a bulk fluidity smaller than $a_w$ 
is described by fixing  $\partial_z a|_{z =0}< 0$, and following the same procedure.

\item{\it What is the limit of existence for the fluidised bulk solution $\s_i$? }

$\s_i (H,a_w)$ was defined in section~\ref{stress_H} as the stress beyond which a profile 
with a frozen bulk can not be observed.

For $H \sim \infty$,  $\s_i(\infty,a_w)$ is defined by $W_{\sigma_i}(a_w)=0$. With the 
mechanical analogy, the mechanical energy $E_m$ is set to $0$, that is the particle leaves 
its position $a_w$, with a velocity $\partial_z a|_{z=0} \sim 0$, and spends an infinite time 
at the position $a_m=0$.

For $\s > \s_i,W_{\sigma}(a_w)>0$, and for $a<a_w ,W_{\sigma}(a)< W_{\sigma}(a_w)$. 
With the mechanical analogy, the particle can not leave the position $a_w$ with a negative 
velocity and come back to it after a time $H$.
There is no solution with a frozen bulk.

For a finite $H$, a frozen bulk requires $E_m \lesssim 0$, therefore $\s_i(H,a_w)$ is slightly 
smaller than $ \s_i(\infty,a_w)$, by an amount that scales as $H^{-3}$.

The situation is symmetric for the definition of $\s_d(H,a_w)$, which is found slightly larger 
than  $\s_d(\infty,a_w)$.

\item{\it How can one calculate the fluidity profile for a solution with several bands?}

The mechanical energy $E_m$ is now decreased from $Min(W_{\Sigma}(0),W_{\Sigma}(a_2))$, 
to $W_{\Sigma}(a_w)$ to find banded stationary solutions : a frozen bulk solution with thickness 
$h_1$ and, a fluidised bulk solution with thickness $h_2$ (both with the same energy $E_m$) 
such that $h_1+ h_2 =H$ form a single interface solution. The same process is derived to find 
multibanded solutions. 
Solutions of figure~\ref{zoom} were calculated numerically with this method. \end{itemize}

\newpage
%%%%%%%%%%%%%%%%%%%%%%%%%%%%%%%%%%%%%%%%%%%%%%%%

 \end{document}